\documentclass[conference,a4paper]{IEEEtran}
\usepackage{amsmath,amssymb,amsfonts}
\usepackage{amsthm}
\usepackage{algorithm}
\usepackage{algorithmic}
\usepackage{array}
\usepackage{booktabs}
\usepackage{cite}
\usepackage{multirow}
\usepackage{tikz}
\usepackage{url}
\usepackage{graphicx}
\usepackage{xcolor}
\usepackage{eso-pic}

\newcommand{\ArchiveOrgTime}{0.494}
\newcommand{\ArchiveCompactTime}{0.497}
\newcommand{\ArchiveTimeChange}{-24.5\%}
\newcommand{\ArchiveChangeQOne}{-54.9\%}
\newcommand{\ArchiveChangeQThree}{-2.0\%}
\newcommand{\ArchiveLowerCount}{95}
\newcommand{\ArchiveCompactSpan}{6.6\%}
\newcommand{\ArchiveCompactSpanQOne}{2.8\%}
\newcommand{\ArchiveCompactSpanQThree}{14.6\%}
\newcommand{\ArchiveOrgSpan}{2.4\%}
\newcommand{\ArchiveOrgSpanQOne}{1.4\%}
\newcommand{\ArchiveOrgSpanQThree}{3.4\%}

\makeatletter
\let\CameraReadyFloatBox\float@makebox
\renewcommand{\float@makebox}[1]{\color@vbox\normalcolor\CameraReadyFloatBox{#1}\color@endbox}
\makeatother

\newtheorem{proposition}{Proposition}

\begin{document}

\title{Compact SAT and MaxSAT Encodings for Business-to-Business Meeting Scheduling with Idle-Time Balancing}

\author{
\IEEEauthorblockN{Long Duc Nguyen, Tuyen Van Kieu, and Khanh Van To}
\IEEEauthorblockA{
Faculty of Information Technology\\
VNU University of Engineering and Technology, Hanoi, Vietnam\\
\{24021555,tuyenkv,khanhtv\}@vnu.edu.vn} }

\maketitle

\AddToShipoutPictureFG*{%
  \AtPageLowerLeft{%
    \put(\LenToUnit{\dimexpr(\paperwidth-\textwidth)/2\relax},\LenToUnit{16mm}){%
      \parbox[b]{\textwidth}{\normalfont\fontsize{7}{8}\selectfont
        \copyright~2026 IEEE. Personal use of this material is permitted.
        Permission from IEEE must be obtained for all other uses, in any current
        or future media, including reprinting/republishing this material for
        advertising or promotional purposes, creating new collective works,
        for resale or redistribution to servers or lists, or reuse of any
        copyrighted component of this work in other works.}%
    }%
  }%
}

\begin{abstract}
Business-to-business meeting scheduling assigns pairwise meetings to slots and locations under availability, capacity, conflict, and precedence constraints. The published Boolean formulation retains removable assignments and uses pairwise precedence clauses. We combine solution-preserving domain filtering, sparse shared-suffix precedence encoding, and an objective minimizing the range of participants' internal idle-slot totals. Experiments on 126 official and 100 higher-density derived instances separate domain filtering, precedence representation, transitive relations, and optimization method. Under the same objective, the compact model achieves median paired reductions of 40.3\% in clauses and 55.9\% in peak memory relative to an adapted published MaxSAT formulation. Domain filtering alone reduces assignment variables by 24.1\% and clauses by 16.2\%. Suffix sharing gives an additional clause reduction of 0.5\%--1.0\% on official precedence instances, reaching 5.5\% at the highest density in derived instances that retain a known feasible schedule. Total-time changes are small and mixed. Internal idle distinguishes short interruptions from longer waits; aggregate idle complements the range. In the original one-run comparison, all three SAT and MaxSAT methods solve every official instance with lower median total times than Gurobi.
\end{abstract}

\begin{IEEEkeywords}
business-to-business meeting scheduling, SAT, domain filtering, precedence constraints, idle-time balancing
\end{IEEEkeywords}

\section{Introduction}
\label{sec:introduction}

Business-to-business (B2B) meeting scheduling assigns requested pairwise meetings to time slots and locations under participant non-overlap, location capacity, availability, session, fixed-slot, forbidden-slot, and precedence constraints~\cite{pesant2015,bofill2022}.  Constraint programming (CP), mixed-integer programming (MIP), propositional satisfiability (SAT), and maximum satisfiability (MaxSAT) formulations have been studied~\cite{bofill2015,bofill2019}.  MaxSAT was reported as the strongest evaluated approach on the extended benchmarks~\cite{bofill2022}.  The published Boolean formulation uses a full meeting--slot matrix and pairwise precedence clauses that can be quadratic in the two meeting-domain sizes~\cite{bofill2022}.  We examine domain reduction and precedence representation as separate modeling choices.

We integrate precedence-bound propagation, per-participant matching~\cite{regin1994}, and saturated-slot propagation before conjunctive normal form (CNF) generation.  This solution-preserving preprocessing removes unsupported assignments.  For precedence, we build on sequential and parametric cardinality encodings~\cite{sinz2005,abio2013}, shared staircase counters, and cumulative precedence constructions~\cite{truong2025,vankieu2025compact}.  The resulting sparse shared-suffix encoding, denoted by \mbox{\textnormal{\textsc{SparseSuffix}}} in tables, materializes only selected predecessor-domain cut points and reuses variables for equal cuts.  The experiments also compare direct precedence arcs with their transitive closure.

The published B2B objective minimizes total break groups and may bound differences in participant break counts~\cite{bofill2022}. Motivated by norm-based load balancing in the Balanced Academic Curriculum Problem (BACP)~\cite{chiarandini2012}, we minimize the range of internal idle-slot totals among participants with at least two meetings. Aggregate idle measures total internal waiting.

This study makes three contributions:
\begin{itemize}
    \item Solution-preserving preprocessing combines established filters; Proposition~\ref{prop:domain-preservation} proves correctness, and a full-versus-reduced comparison measures formula size, memory, and time.
    \item An exact shared-suffix encoding is equivalent to pairwise clauses (Proposition~\ref{prop:suffix-equivalence}); a factorized study separates suffix sharing from domain reduction.
    \item A B2B idle-time range objective has an exact threshold encoding based on first and last occupied slots, using linearly many auxiliary variables.
\end{itemize}

We study domain reduction (RQ1), filtering and precedence choices (RQ2), SAT and MaxSAT optimization (RQ3), and comparison with Gurobi, CPLEX, and CP Optimizer (RQ4).

\section{Background and Related Work}
\label{sec:related}

Early matchmaking used answer set programming~\cite{gebser2013}. Pesant et al.\ compared CP and MIP for minimizing break groups~\cite{pesant2015}. Bofill et al.\ introduced partial MaxSAT~\cite{bofill2015}, studied implied constraints~\cite{bofill2019,bofill2022impact}, and compared CP, MIP, and MaxSAT on extended benchmarks~\cite{bofill2022}. We use their problem definition and benchmarks. Their MaxSAT formulation uses full meeting--slot domains and pairwise direct-precedence clauses. Our evaluation separates domain reduction, filtering relation, precedence representation, and encoded relation set.

Sequential counters give linear-size at-most-one encodings, while parametric cardinality encodings trade auxiliary variables against clauses~\cite{sinz2005,abio2013}. Pseudo-Boolean encodings exploit implied at-most-one relations~\cite{bofill2022pbamo}; precedence paths identify such relations in resource-constrained scheduling~\cite{bofill2020}. Shared staircase counters and cumulative precedence encodings reuse auxiliary states~\cite{truong2025,vankieu2025compact}. SparseSuffix adapts this sharing to reduced B2B domains, creating only queried cut points and sharing equal cuts. Section~\ref{sec:precedence-encoding} proves equivalence to pairwise clauses and separates suffix sharing from the encoded relation set. For curriculum-load balancing, Chiarandini et al.\ use norm-based deviations from an ideal distribution~\cite{chiarandini2012}.

\section{Problem Definition and Idle-Time Range Objective}
\label{sec:problem}

Let $M$, $P$, and $T=\{1,\ldots,h\}$ denote the sets of meetings, participants, and ordered time slots.  Each meeting $m\in M$ involves exactly two participants, denoted by $\pi(m)\subseteq P$ with $|\pi(m)|=2$, and has an initial slot domain $D_m^0\subseteq T$ determined by its requested session, fixed-slot requirement, and the two participants' forbidden-slot restrictions. Let $K$ be the number of interchangeable locations and let $E\subseteq M\times M$ be the set of directed input precedence arcs.  A schedule $S:M\rightarrow T$ is feasible when $S(m)\in D_m^0$ and
\begin{align}
S(m)&\ne S(m') &&
  \substack{\forall m,m'\in M,\ m\ne m':\\
  \pi(m)\cap\pi(m')\ne\emptyset}, \label{eq:participant}\\
\left|\{m\in M:S(m)=t\}\right|&\le K &&
  \forall t\in T, \label{eq:capacity}\\
S(i)&<S(j) &&
  \forall(i,j)\in E. \label{eq:precedence}
\end{align}

Each meeting occupies one time slot.  Since every meeting can use any of the $K$ interchangeable locations, the capacity condition is sufficient to extend a feasible slot schedule to a location assignment. Rooms with different capacities or equipment would require assigning each meeting to a suitable room as well as a time slot. For $M_p=\{m\in M:p\in\pi(m)\}$ and $|M_p|\geq1$, define the first and last occupied slots by
\begin{equation*}
F_p(S)=\min_{m\in M_p}S(m),\qquad
L_p(S)=\max_{m\in M_p}S(m).
\end{equation*}
Participant non-overlap makes these occupied slots distinct, so the number of unoccupied slots strictly inside this span is
\begin{equation}
I_p(S)=
\begin{cases}
L_p(S)-F_p(S)+1-|M_p|, & |M_p|\ge 2,\\
0, & |M_p|\le 1.
\end{cases}
\label{eq:internal-idle}
\end{equation}
Leading and trailing free slots are excluded.  A break group is a maximal contiguous block of unoccupied slots between two occupied slots \cite{bofill2022}.  Meetings in slots $\{1,3\}$ and $\{1,7\}$ each create one break group, whereas $I_p$ assigns them one and five internal idle slots, respectively.  The established objective minimizes the total number of break groups and, in its main benchmark setting, constrains the difference between participant break counts by $d=2$~\cite{bofill2022}.  Here, the range of participant internal idle-slot counts expresses the desired balance directly. The break-group objective is suitable when fewer interruptions are the priority. The range objective is suitable when participants should have similar waiting durations.

Internal idle slots arise only for participants with at least two meetings. We therefore define $P^\star=\{p\in P:|M_p|\ge2\}$.  For a feasible schedule $S$ and $|P^\star|\ge2$, define
\begin{equation}
\Delta_I(S)=\max_{p\in P^\star}I_p(S)-\min_{p\in P^\star}I_p(S).
\label{eq:idle-range}
\end{equation}
We minimize $\Delta_I(S)$ and report, rather than minimize, aggregate idle $I_\Sigma(S)=\sum_{p\in P}I_p(S)$. Equal ranges can have different sums: $[0,0]$ and $[5,5]$ both have range zero, but totals 0 and 10.

\section{Compact SAT Encoding}
\label{sec:encoding}

The encoding first filters the meeting domains, creates assignment variables over the selected domains, encodes feasibility and precedence, and derives the idle-time objective.  The precedence relation used for domain filtering is selected independently from the relation encoded in CNF.

\subsection{Domain Filtering and Assignment Variables}

Starting from $D_m^0$ in Section~\ref{sec:problem}, define $\mathcal R_E=\{(i,j,1):(i,j)\in E\}$ and $\mathcal R_{E^\star}=\{(i,j,\ell_{ij}):i\rightsquigarrow j\}$, where $\ell_{ij}$ is the longest $i$--$j$ path length for acyclic $E$. Algorithm~\ref{alg:domains} selects $\mathcal R_f$ as $\mathcal R_E$ (\mbox{\textsc{Filter-}$E$}) or $\mathcal R_{E^\star}$ (\mbox{\textsc{Filter-}$E^\star$}) and repeats three filters. For $(i,j,d)\in\mathcal R_f$, remove $t\in D_i$ if $t+d>\max D_j$ and $u\in D_j$ if $\min D_i+d>u$. Per-participant matching removes assignments absent from every conflict-free assignment of that participant's meetings~\cite{regin1994}. If $K$ meetings have singleton domain $\{t\}$, remove $t$ from all others. A cycle, empty domain, or overloaded slot proves infeasibility; otherwise iteration yields $D_m$. These necessary conditions preserve every feasible schedule.

\begin{algorithm}[t]
\caption{Solution-Preserving Domain Filtering}
\label{alg:domains}
\small
\begin{algorithmic}[1]
\REQUIRE domains $\{D_m^0\}$, arcs $E$, meetings $\{M_p\}$, capacity $K$, filter \mbox{\textnormal{\textsc{Filter-}}\ensuremath{E}}/\mbox{\textnormal{\textsc{Filter-}}\ensuremath{E^\star}}
\STATE reject a directed cycle in $E$; set $\mathcal R_f\leftarrow\mathcal R_E$ or $\mathcal R_{E^\star}$; $D_m\leftarrow D_m^0$
\REPEAT
  \STATE apply endpoint-bound filters for every $(i,j,d)\in\mathcal R_f$
  \STATE remove assignments absent from any complete matching of each $M_p$
  \STATE propagate slots forced to $K$ meetings; reject empty domain or overloaded slot
\UNTIL{no domain changes}
\RETURN $\{D_m\}$
\end{algorithmic}
\end{algorithm}

\begin{proposition}
\label{prop:domain-preservation}
For either filtering choice, Algorithm~\ref{alg:domains} terminates, preserves every feasible schedule, and reports \textup{\textsc{Infeasible}} only for an infeasible instance.
\end{proposition}
\emph{Proof sketch.} Every nonfinal iteration deletes a value from finite domains.  The filters remove only values that violate a precedence bound, lack support in a complete per-participant matching, or use capacity already forced to other meetings. Each reported contradiction is necessary for feasibility.  With $N_0=\sum_m|D_m^0|$ and $r_p=|M_p|$, at most $N_0$ iterations can delete values.  The implemented support tests give the conservative bound $O(|M|^3+(N_0+1)(|\mathcal R_f|h+|M|h+\sum_p r_p^3h^2))$. The $O(|M|^3)$ term covers computing the transitive closure and longest-path distances in the precedence graph. Each filtering pass takes $O(|\mathcal R_f|h)$ for precedence bounds and $O(|M|h)$ for saturated slots. For participant $p$, at most $r_ph$ meeting--slot choices are checked, each by finding a matching in $O(r_p^2h)$ time, giving $O(r_p^3h^2)$ per pass. The factor $N_0+1$ includes the final pass with no deletions.

Let $A_m$ be the domain represented by meeting-slot assignment variables.  The \mbox{\textsc{Full}} domain representation sets $A_m=T$, creates $|M||T|$ variables, and excludes $T\setminus D_m^0$ by unit clauses.  The \mbox{\textsc{Reduced}} domain representation sets $A_m=D_m$ and creates $\sum_m|D_m|$ variables $x_{m,t}$ for $t\in A_m$.  Both representations use the same encoding methods and configuration choices; they differ in the active meeting--slot domains supplied to these encoders. Proposition~\ref{prop:domain-preservation} therefore gives a one-to-one correspondence between their feasible schedules.

\subsection{Feasibility Constraints}

For $t\in A_m$, let $x_{m,t}$ be true exactly when meeting $m$ is assigned to slot $t$.  The formulation imposes
\begin{align}
\textstyle\sum_{t\in A_m}x_{m,t}=1\;\forall m, &\quad
\textstyle\sum_{\substack{m\in M_p\\t\in A_m}}\!x_{m,t}\leq1\;\forall p\in P,\ t\in T, \notag\\
&\quad\textstyle\sum_{\substack{m\in M\\t\in A_m}}\!x_{m,t}\leq K\;\forall t\in T.
\label{eq:core-cnf}
\end{align}
The implementation uses a recursive commander encoding for exactly-one constraints~\cite{klieber2007} and pairwise participant-conflict clauses.  The fixed \textsc{IC12+} configuration combines participant-based clusters, sequential-counter capacity~\cite{sinz2005,abio2013}, and the implied constraints on occupied slots and the even number of busy participants studied in prior B2B MaxSAT work~\cite{bofill2019,bofill2022impact}.

\subsection{Precedence Encodings and Transitive Closure}
\label{sec:precedence-encoding}

The CNF relation factor independently selects $\mathcal R_g\in\{\mathcal R_E,\mathcal R_{E^\star}\}$. \mbox{\textnormal{\textsc{Direct-}}\ensuremath{E}} uses $\mathcal R_E$, while \mbox{\textnormal{\textsc{Closure-}}\ensuremath{E^\star}} uses $\mathcal R_{E^\star}$.

\begin{proposition}
\label{prop:closure-equivalence}
For acyclic $E$, \mbox{\textnormal{\textsc{Direct-}}\ensuremath{E}} and \mbox{\textnormal{\textsc{Closure-}}\ensuremath{E^\star}} represent exactly the same meeting schedules.
\end{proposition}
\emph{Proof sketch.} Summing inequalities along a longest $i$--$j$ path gives $S(i)+\ell_{ij}\leq S(j)$, so direct constraints imply closure constraints. Conversely, each direct arc's closure constraint implies its original inequality. Holding the filtering relation fixed therefore preserves feasibility and assignment variables, isolating the effect of $\mathcal R_g$.

For a selected relation $(i,j,d)\in\mathcal R_g$, the \mbox{\textnormal{\textsc{Pairwise}}} encoding adds
\begin{equation}
\neg x_{i,t}\lor\neg x_{j,u}
\quad
\forall t\in A_i,\ u\in A_j:\ t+d>u.
\label{eq:pairwise-precedence}
\end{equation}
It can generate $O(|A_i||A_j|)$ binary clauses for one relation.

For sorted $A_i=(a_{i,0},\ldots,a_{i,n_i-1})$, each successor assignment determines the first predecessor-domain index whose slot would be too late. \mbox{\textnormal{\textsc{SparseSuffix}}} creates variables only for the distinct interior cut indices that occur.  If these indices are $c_1<\cdots<c_s$, set $c_{s+1}=n_i$ and $q_{i,n_i}=\bot$:
\begin{align}
c(i,j,u)&=\min\{r\in\{0,\ldots,n_i-1\}:a_{i,r}>u-d\}, \notag\\
q_{i,c_\ell}&\leftrightarrow
\left(\bigvee_{r=c_\ell}^{c_{\ell+1}-1}x_{i,a_{i,r}}\right)
\lor q_{i,c_{\ell+1}}, \label{eq:sparse-suffix}\\
x_{j,u}&\rightarrow\neg q_{i,c(i,j,u)}.
\label{eq:suffix-link}
\end{align}
Take $c=n_i$ if the set is empty.  For $c=0$ the last line becomes $\neg x_{j,u}$.  For $c=n_i$ it is omitted.  The same $(i,c)$ suffix is reused whenever two relations produce the same predecessor-domain cut.

\begin{proposition}
\label{prop:suffix-equivalence}
The \mbox{\textnormal{\textsc{Pairwise}}} and \mbox{\textnormal{\textsc{SparseSuffix}}} encodings represent the same meeting-slot assignments for either relation set.
\end{proposition}
\emph{Proof sketch.} The recurrence makes $q_{i,c}$ the exact disjunction of assignments with index at least $c$.  The link excludes exactly the pairs with $t+d>u$.  With $C_i$ queried cut indices, construction uses at most $C_i$ variables, $O(|A_i|+C_i)$ defining clauses, and at most one link clause per successor value and relation.

\subsection{Idle-Time Thresholds from the Schedule Span}

Let $y_{p,t}\leftrightarrow\bigvee_{m\in M_p:t\in A_m}x_{m,t}$ indicate that participant $p$ is occupied at $t$.  For $p\in P^\star$, define two running occupancy indicators.  $P_{p,t}$ records whether $p$ is occupied at or before $t$, and $R_{p,t}$ records whether $p$ is occupied at or after $t$: $P_{p,t}\leftrightarrow P_{p,t-1}\lor y_{p,t}$ and $R_{p,t}\leftrightarrow y_{p,t}\lor R_{p,t+1}$, with $P_{p,0}=\bot$ and $R_{p,s}=\bot$ for every $s>h$.  Let $f_{p,t}\leftrightarrow y_{p,t}\land\neg P_{p,t-1}$ identify the first occupied slot.  Every feasible schedule has exactly one true $f_{p,t}$ for each $p\in P^\star$.

Let $\theta_{p,k}$ denote $I_p(S)\geq k$, and let $A_p=\bigcup_{m\in M_p}A_m$.  For nonempty $A_p$, define the valid upper bound $U_p=\max\{0,\max A_p-\min A_p+1-|M_p|\}$; set $U_p=0$ when $A_p=\emptyset$.  Equation~\eqref{eq:internal-idle} gives
\begin{equation}
f_{p,t}\ \rightarrow\
\left(\theta_{p,k}\leftrightarrow R_{p,t+|M_p|+k-1}\right)
\qquad \forall t\in T,\ 1\le k\le U_p.
\label{eq:span-threshold}
\end{equation}
Thus $I_p(S)\ge k$ holds exactly when the last meeting is far enough from the first to contain $|M_p|$ occupied and $k$ unoccupied slots.  For example, with $|M_p|=2$, a first meeting at $t=3$, and $k=1$, the condition is $R_{p,5}$: the last meeting must be at slot 5 or later to leave at least one internal idle slot. Figure~\ref{fig:encoding-ideas} illustrates a predecessor-domain cut and the idle-time test based on the first and last occupied slots.

\begin{figure}[t]
\centering
\begin{tikzpicture}[x=0.47cm,y=0.46cm,font=\scriptsize]
\begin{scope}
  \node[align=center] at (2.5,4.75) {(a) Predecessor-domain cut};
  \node[anchor=east] at (0.8,3.30) {$A_i$};
  \foreach \x/\v in {1/1,2/3,3/4,4/7} {
    \draw (\x,3.00) rectangle +(0.70,0.56);
    \node at (\x+0.35,3.28) {$\v$};
  }
  \draw[black,dashed,thick] (2.82,2.85)--(2.82,3.62);
  \node[black,anchor=south] at (2.82,3.62) {$c=2$};
  \draw[black,thick] (2.88,2.76) -- (4.70,2.76);
  \node[black,anchor=north] at (2.6,2.65)
    {$q_{i,2}=x_{i,4}\lor x_{i,7}$};
  \node[anchor=west] at (5.0,3.28) {$x_{j,4}$};
  \node at (2.6,1.70) {$\neg x_{j,4}\lor\neg q_{i,2}$};
\end{scope}
\begin{scope}[shift={(8.3,0)}]
  \node[anchor=center] at (2.5,4.75) {(b) Idle-time threshold};
  \node[anchor=east] at (0.8,2.87) {slot};
  \foreach \x/\v in {1/3,2/4,3/5,4/6} {
    \draw (\x,2.57) rectangle +(0.70,0.56);
    \node at (\x+0.35,2.85) {$\v$};
  }
  \fill[black] (1.12,2.70) rectangle +(0.46,0.30);
  \fill[black] (3.12,2.70) rectangle +(0.46,0.30);
  \node[anchor=south] at (1.35,3.15) {$f_{p,3}$};
  \node[anchor=south] at (3.35,3.15) {$R_{p,5}$};
  \draw[<->,black,thick] (1.35,2.38)--(3.35,2.38);
  \node[black,anchor=north] at (2.5,1.70)
    {$|M_p|=2,\ I_p\ge1$};
\end{scope}
\end{tikzpicture}
\caption{(a) Predecessor-domain cut for SparseSuffix; (b) span-based idle-time threshold.}
\label{fig:encoding-ideas}
\end{figure}

When $|P^\star|\leq1$, no objective literals are required.  Otherwise, let $U=\max_{p\in P^\star}U_p$ and treat $\theta_{p,k}$ as false for $k>U_p$. For $1\leq k\leq U$, introduce objective literals:
\begin{align}
\Theta_k^{\max}&\leftrightarrow\bigvee_{p\in P^\star}\theta_{p,k}, &
\Theta_k^{\min}&\leftrightarrow\bigwedge_{p\in P^\star}\theta_{p,k}, \notag\\
g_k&\leftrightarrow \Theta_k^{\max}\land\neg \Theta_k^{\min}, &
\sum_{k=1}^{U}g_k&=\Delta_I(S).
\label{eq:range-sum}
\end{align}
Here $g_k$ is true exactly when $\min_{p\in P^\star}I_p(S)<k\leq\max_{p\in P^\star}I_p(S)$. Consequently, the true $g_k$ literals correspond to the integer levels between the minimum and maximum idle counts, which proves the final equality in Eq.~\eqref{eq:range-sum}.  The span and range construction introduces $O(|P||T|+\sum_{p\in P^\star}U_p)$ auxiliary variables and $O(|M||T|+|P||T|+ \sum_{p\in P^\star}|A_p|U_p)$ clauses.  The experiments report the generated counts.

\section{Optimization Approaches}
\label{sec:optimization}

Let $F$ be the base CNF from Section~\ref{sec:encoding}, including the exact definitions of $g_1,\ldots,g_U$ in Eq.~\eqref{eq:range-sum}. UWrMaxSAT treats $F$ as hard and assigns unit weight to each soft clause $\neg g_k$, so a schedule $S$ has cost $\sum_{k=1}^{U}g_k=\Delta_I(S)$.  \textsc{Non-Incremental SAT} and \textsc{Incremental SAT} first solve $F$, either proving infeasibility or obtaining an upper bound $B_0$; $B_0=0$ is optimal.  Otherwise, both binary-search $B\in[0,B_0-1]$ by testing $F\land\operatorname{CNF}(\sum_{k=1}^{U}g_k\leq B)$, whose feasibility is monotone in $B$.  \textsc{Non-Incremental SAT} builds a fresh CaDiCaL instance and bound encoding at each iteration, with a sequential counter for $B>0$~\cite{sinz2005}.  \textsc{Incremental SAT} adds one totalizer to one solver and selects bounds by assumptions, retaining learned clauses.

The commercial baselines use \mbox{\textsc{Reduced}} domains from \mbox{\textnormal{\textsc{Filter-}}\ensuremath{E^\star}} and, on precedence instances, distance-labelled \mbox{\textnormal{\textsc{Closure-}}\ensuremath{E^\star}} relations.  Gurobi and CPLEX share a solver-independent MIP: meeting--slot assignments define occupancy, prefix and suffix recurrences compute $I_p$, and integer min/max variables define $\Delta_I$.  CP Optimizer uses meeting-time variables over the same domains, global all-different and count constraints, and $\tau_i+d\leq\tau_j$ for each closure relation.  Native min/max expressions compute $\Delta_I$.  Both formulations enforce the same feasibility conditions and minimize $\Delta_I$ over $P^\star$.

\section{Experimental Methodology}
\label{sec:experiments}

The official \emph{All-126} benchmarks~\cite{bofill2022}\footnote{\url{https://imae.udg.edu/Recerca/LAI/}} contain 20 original, 26 forbidden, 40 fixed, 20 \texttt{prec15}, and 20 \texttt{prec25} instances. Instances contain 125--302 meetings, 42--78 participants, and 8--22 slots. The precedence subset, \emph{Prec-40}, has 14--126 direct arcs (median 48). Results use All-126 unless stated otherwise. \emph{Stress-100} contains five variants, $\gamma\in\{30,35,40,50,60\}$, of each original instance. Each participant with $d$ meetings receives exactly $\lfloor\gamma d/100\rfloor$ incoming arcs. Nested edge sets follow one fixed feasible schedule per source instance, ensuring acyclicity and retaining that schedule at all five densities without changing problem size. This tests sensitivity to precedence density within these variants. Performance on other dense precedence patterns remains to be evaluated.

RQ1 pairs the \mbox{\textsc{Full}} and \mbox{\textsc{Reduced}} domain representations on All-126 for formula size and total time while fixing \mbox{\textnormal{\textsc{Filter-}}\ensuremath{E^\star}}, \mbox{\textnormal{\textsc{Pairwise}}}, \mbox{\textnormal{\textsc{Direct-}}\ensuremath{E}}, UWrMaxSAT, and \textsc{IC12+}.  RQ2 evaluates all $2\times2\times2\times3=24$ combinations of \mbox{\textnormal{\textsc{Filter-}}\ensuremath{E}} versus \mbox{\textnormal{\textsc{Filter-}}\ensuremath{E^\star}}, \mbox{\textnormal{\textsc{Pairwise}}} versus \mbox{\textnormal{\textsc{SparseSuffix}}}, \mbox{\textnormal{\textsc{Direct-}}\ensuremath{E}} versus \mbox{\textnormal{\textsc{Closure-}}\ensuremath{E^\star}} (Proposition~\ref{prop:closure-equivalence}), and the three optimization methods.  Each combination uses \mbox{\textsc{Reduced}} domains and is tested on the 140 precedence cases in Prec-40 and Stress-100.  The filter-convergence summary reports both domain-filtering relations at every density.  The precedence-density summary fixes \mbox{\textsc{Reduced}}, \mbox{\textnormal{\textsc{Filter-}}\ensuremath{E^\star}}, and UWrMaxSAT.  RQ3 compares the three SAT and MaxSAT optimization methods on All-126. RQ4 adds Gurobi MIP, CPLEX MIP, and CP Optimizer, and compares all six exact methods on All-126.

The compact model reported in the overall comparison uses \mbox{\textsc{Reduced}} domains obtained with \mbox{\textnormal{\textsc{Filter-}}\ensuremath{E^\star}}.  On precedence instances, it combines \mbox{\textnormal{\textsc{SparseSuffix}}} with \mbox{\textnormal{\textsc{Closure-}}\ensuremath{E^\star}}.  Otherwise, it uses \mbox{\textnormal{\textsc{Pairwise}}} with \mbox{\textnormal{\textsc{Direct-}}\ensuremath{E}}. \mbox{\textsc{ORG-MaxSAT}} is an equation-level adaptation of the published MaxSAT formulation~\cite{bofill2022}.  It retains full meeting--slot domains, \mbox{\textnormal{\textsc{Pairwise}}} with \mbox{\textnormal{\textsc{Direct-}}\ensuremath{E}} precedence, a slot-based cardinality encoding of idle time, and \textsc{IC12+}.  It uses the common $\Delta_I$ objective without the original fairness term.  In the one-run comparison, the baseline and compact models use the same UWrMaxSAT binary, machine, and time limit.

The original comparisons use one run per configuration--instance pair.  Runs used one thread, random seed 0 where supported, and a time limit of 7{,}200~s on the same 4-core GCP virtual machine with an Intel Xeon Platinum 8581C processor and 15.6~GB of RAM.  The software environment comprised Linux 6.8, Python 3.10.12, PySAT 1.9.dev7, CaDiCaL 1.5.3, UWrMaxSAT, Gurobi 13.0.2, CPLEX 22.2.0.0, and CP Optimizer through DOcplex 2.32.264. The commercial solvers represent the MIP and CP paradigms used in prior B2B comparisons~\cite{pesant2015,bofill2022}.  Total time includes input parsing, encoding or model construction, solving, decoding, and validation.  Peak memory is the largest observed combined physical-memory use of the benchmark process and its solver subprocesses. Terminal statuses and objectives were cross-checked across matched methods. RQ2 additionally records model-build time from before parsing through CNF construction; for SAT and MaxSAT, it includes domain filtering but excludes solving.

We also compare both models using three runs per All-126 instance under the same experimental setup. Paired comparisons use each model's three-run median for each instance. Variation is the difference between longest and shortest times divided by their median. Relative changes are medians of paired per-instance $(\text{candidate}-\text{reference})/\text{reference}$ values; negative means reduction. Tables~\ref{tab:compactness} and~\ref{tab:filter-graph} give medians across instances. Table~\ref{tab:filter-graph} first takes each instance's median model-build time over 12 matched encoding, relation-set, and optimizer combinations. Fixpoint counts come from deterministic replay of the filtering loop. Table~\ref{tab:compactness} bolds smaller formula-size and memory values. ``Solved'' includes proven-optimal and proven-infeasible outcomes.

{\raggedright Code and data are available at \url{https://github.com/Longnguyen1555/Business-Meeting-Scheduling-Problem}. \par}

\section{Results and Discussion}
\label{sec:results}

\subsection{Formula Size and Domain Reduction (RQ1)}

\begin{table}[t]
\caption{All-126 Medians and Median Paired Changes. One Run Unless Marked Three; Changes Calculated per Instance.}
\label{tab:compactness}
\centering
\scriptsize
\setlength{\tabcolsep}{2.7pt}
\begin{tabular}{@{}lccc@{}}
\toprule
\emph{Overall} & \mbox{\textsc{ORG-MaxSAT}} & Compact & Paired change $\downarrow$ \\
\midrule
Total variables & 27859.0 &
  \textbf{26727.5} & \ensuremath{-}17.6\% \\
Total clauses & 93183.5 &
  \textbf{72487.0} & \ensuremath{-}40.3\% \\
Peak memory (MB) & 160.9 &
  \textbf{68.9} & \ensuremath{-}55.9\% \\
Total time (s) & 0.467 &
  0.432 & \ensuremath{-}14.0\% \\
Total time (s), three runs & \ArchiveOrgTime &
  \ArchiveCompactTime & \ArchiveTimeChange \\
\addlinespace[1pt]
\emph{Domain reduction} & \mbox{\textsc{Full}} & \mbox{\textsc{Reduced}} & Paired change $\downarrow$ \\
\midrule
Assignment variables & 2370.0 &
  \textbf{2059.5} & \ensuremath{-}24.1\% \\
Total clauses & 76396.0 &
  \textbf{72649.0} & \ensuremath{-}16.2\% \\
Peak memory (MB) & 71.9 &
  \textbf{69.1} & \ensuremath{-}7.4\% \\
Total time (s) & 0.529 &
  0.438 & \ensuremath{-}12.3\% \\
\bottomrule
\end{tabular}
\end{table}
Table~\ref{tab:compactness} compares the overall configurations and isolates domain reduction on All-126. The compact configuration reduces clauses and peak memory on every instance, with median paired reductions of 40.3\% and 55.9\%. Its one-run median paired time reduction is 14.0\%, with lower times on 88/126 instances. The three-run comparison gives a median paired time change of \ArchiveTimeChange{} [IQR \ArchiveChangeQOne{} to \ArchiveChangeQThree{}], with a lower median total time on \ArchiveLowerCount/126 instances. The median variation across three runs is \ArchiveCompactSpan{} [IQR \ArchiveCompactSpanQOne{}--\ArchiveCompactSpanQThree{}] for the compact model and \ArchiveOrgSpan{} [IQR \ArchiveOrgSpanQOne{}--\ArchiveOrgSpanQThree{}] for \mbox{\textsc{ORG-MaxSAT}}.

With all other components fixed, \mbox{\textsc{Reduced}} decreases the assignment-variable count by 24.1\% and the clause count by 16.2\%.  It also yields a median paired total-time reduction of 12.3\% and has lower total time on 108 of 126 instances.  These comparisons identify domain reduction as the largest measured source of formula-size reduction.  The overall comparison also includes the idle-time and precedence encodings.

\subsection{Domain Filtering and Precedence Representation (RQ2)}

Table~\ref{tab:filter-graph} compares filtering convergence and model-build time on \mbox{\textsc{Reduced}} domains at all seven densities. Each pass applies precedence, matching, and saturated-slot propagation; the count includes the final no-change pass.

\begin{table}[t]
\caption{Median Fixpoint Passes and Model-Build Times under \mbox{\textnormal{\textsc{Filter-}}\ensuremath{E}} and
\mbox{\textnormal{\textsc{Filter-}}\ensuremath{E^\star}}}
\label{tab:filter-graph}
\centering
\scriptsize
\setlength{\tabcolsep}{1.2pt}
\begin{tabular}{@{}lcccccc@{}}
\toprule
& & \multicolumn{2}{c}{Fixpoint passes} &
  \multicolumn{2}{c}{Model-build time (s)} & \\
\cmidrule(lr){3-4}\cmidrule(lr){5-6}
Set & $N$ & \mbox{\textnormal{\textsc{Filter-}}\ensuremath{E}} & \mbox{\textnormal{\textsc{Filter-}}\ensuremath{E^\star}} & \mbox{\textnormal{\textsc{Filter-}}\ensuremath{E}} & \mbox{\textnormal{\textsc{Filter-}}\ensuremath{E^\star}} &
  \shortstack{\mbox{\textnormal{\textsc{Filter-}}\ensuremath{E^\star}} fewer\\passes} \\
\midrule
 \texttt{prec15} & 20 & 3.0 &
  \textbf{2.0} &
  0.274 &
  \textbf{0.206} &
  12/20 \\
 \texttt{prec25} & 20 & 4.0 &
  \textbf{2.0} &
  0.276 &
  \textbf{0.181} &
  12/20 \\
\midrule
 \texttt{prec30} & 20 & 4.0 &
  \textbf{2.0} &
  0.326 &
  \textbf{0.196} &
  20/20 \\
 \texttt{prec35} & 20 & 4.0 &
  \textbf{2.0} &
  0.317 &
  \textbf{0.195} &
  20/20 \\
 \texttt{prec40} & 20 & 5.0 &
  \textbf{2.0} &
  0.330 &
  \textbf{0.229} &
  20/20 \\
 \texttt{prec50} & 20 & 5.0 &
  \textbf{2.5} &
  0.302 &
  \textbf{0.225} &
  18/20 \\
 \texttt{prec60} & 20 & 6.0 &
  \textbf{3.5} &
  0.305 &
  \textbf{0.219} &
  16/20 \\
\bottomrule
\end{tabular}
\end{table}

\begin{figure}[t]
\centering
\includegraphics[width=\columnwidth]{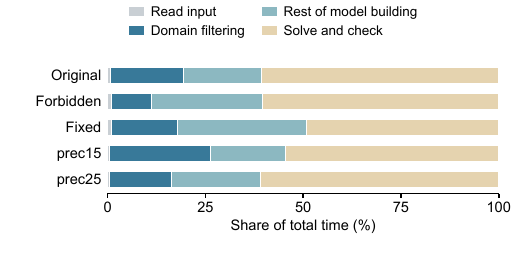}
\caption{Compact model's mean time shares, averaged over three runs per instance, then over instances within each benchmark family.}
\label{fig:phase-breakdown}
\end{figure}

Both filters reach identical reduced domains for all 100 Stress-100 cases and the 30 feasible official precedence cases.  The ten differing official cases are infeasible, and preprocessing terminates as soon as infeasibility is detected. Iterating the direct-arc filter propagates precedence bounds along paths, which explains the identical feasible-domain fixpoints.  Explicit closure propagates the same path bounds in fewer passes: \mbox{\textnormal{\textsc{Filter-}}\ensuremath{E^\star}} uses fewer passes on 118/140 cases and never uses more.  It also has the lower median model-build time at every density; these timing values are descriptive because the two filter campaigns were run in separate batches. Figure~\ref{fig:phase-breakdown} shows time shares for reading input, filtering domains, building the remaining model, and solving and checking the schedule. Each stage time is divided by total time before averaging. Across benchmark families, mean domain-filtering shares range from 10.3\% to 25.8\%. All three filters are timed together. Graph preparation is included in the remaining model-build time.

Table~\ref{tab:precedence} compares precedence encodings with \mbox{\textsc{Reduced}}, \mbox{\textsc{Filter-}$E^\star$}, and UWrMaxSAT. Bold marks the relation set with the larger SparseSuffix reduction per density and metric; ties remain unbolded.
\begin{table}[t]
\caption{Median Paired Changes of \mbox{\textnormal{\textsc{SparseSuffix}}} Relative to \mbox{\textnormal{\textsc{Pairwise}}}
Across Prec-40 and Stress-100}
\label{tab:precedence}
\centering
\scriptsize
\setlength{\tabcolsep}{1.3pt}
\begin{tabular}{@{}lccccc@{}}
\toprule
& & \multicolumn{2}{c}{\mbox{\textnormal{\textsc{Direct-}}\ensuremath{E}} $\downarrow$} &
  \multicolumn{2}{c}{\mbox{\textnormal{\textsc{Closure-}}\ensuremath{E^\star}} $\downarrow$} \\
\cmidrule(lr){3-4}\cmidrule(lr){5-6}
Density & $N$ & Clauses & Total time & Clauses & Total time \\
\midrule
\texttt{prec15} & 20 & \ensuremath{-}0.5\% &
  \textbf{\ensuremath{-}0.1\%} &
  \ensuremath{-}0.5\% &
  0.4\% \\
\texttt{prec25} & 20 & \ensuremath{-}0.7\% &
  \textbf{\ensuremath{-}2.1\%} &
  \textbf{\ensuremath{-}1.5\%} &
  1.7\% \\
\texttt{prec30} & 20 & \ensuremath{-}1.5\% &
  \textbf{\ensuremath{-}2.4\%} &
  \textbf{\ensuremath{-}2.4\%} &
  \ensuremath{-}1.8\% \\
\texttt{prec35} & 20 & \ensuremath{-}1.7\% &
  1.5\% &
  \textbf{\ensuremath{-}2.9\%} &
  \textbf{\ensuremath{-}3.8\%} \\
\texttt{prec40} & 20 & \ensuremath{-}2.0\% &
  \textbf{\ensuremath{-}0.8\%} &
  \textbf{\ensuremath{-}4.1\%} &
  \ensuremath{-}0.5\% \\
\texttt{prec50} & 20 & \ensuremath{-}2.2\% &
  \ensuremath{-}1.5\% &
  \textbf{\ensuremath{-}5.4\%} &
  \textbf{\ensuremath{-}2.5\%} \\
\texttt{prec60} & 20 & \ensuremath{-}1.9\% &
  \ensuremath{-}0.7\% &
  \textbf{\ensuremath{-}5.5\%} &
  \textbf{\ensuremath{-}1.4\%} \\
\midrule
Combined & 140 & \ensuremath{-}1.6\% &
  \textbf{\ensuremath{-}0.9\%} &
  \textbf{\ensuremath{-}2.9\%} & \ensuremath{-}0.7\% \\
\bottomrule
\end{tabular}
\end{table}
On Prec-40, \mbox{\textnormal{\textsc{SparseSuffix}}} increases the total variable count by 1.0\%.  It reduces the total clause count by 0.5\% with \mbox{\textnormal{\textsc{Direct-}}\ensuremath{E}} and by 1.0\% with \mbox{\textnormal{\textsc{Closure-}}\ensuremath{E^\star}}.  Before suffix sharing, domain reduction removes 49.5\% and 55.4\% of pairwise precedence clauses under \mbox{\textnormal{\textsc{Direct-}}\ensuremath{E}} and \mbox{\textnormal{\textsc{Closure-}}\ensuremath{E^\star}}, respectively.  The remaining precedence clauses account for only 2.4\% and 2.7\% of the entire CNF formula. Across all evaluated densities, \mbox{\textnormal{\textsc{SparseSuffix}}} reduces the median clause count relative to \mbox{\textnormal{\textsc{Pairwise}}}.  \mbox{\textnormal{\textsc{Closure-}}\ensuremath{E^\star}} yields the larger reduction from \texttt{prec25} through \texttt{prec60}, reaching 5.5\% at \texttt{prec60} in Stress-100.  The combined median paired total-time changes are \ensuremath{-}0.9\% with \mbox{\textnormal{\textsc{Direct-}}\ensuremath{E}} and \ensuremath{-}0.7\% with \mbox{\textnormal{\textsc{Closure-}}\ensuremath{E^\star}}; their direction varies across individual densities.

\subsection{Optimization-Method Comparison (RQ3--RQ4)}

All three SAT and MaxSAT methods agree on every All-126 status and feasible optimum. Of 39 returned zero-range schedules, 12 have $I_\Sigma=116$, with two idle slots for each participant in $P^\star$. Table~\ref{tab:idle-quality} summarizes the 89 feasible UWrMaxSAT schedules: overall median 47 [IQR 0--116]. Table~\ref{tab:cross-paradigm} uses all 126 one-run total times, including timeouts. Penalized average runtime (PAR-2) assigns each timeout 14{,}400~s before taking the mean. Bold marks the best median and PAR-2 among methods solving all instances.

\begin{table}[t]
\caption{Aggregate Idle in the 89 Range-Optimal UWrMaxSAT Schedules}
\label{tab:idle-quality}
\footnotesize
\setlength{\tabcolsep}{3pt}
\newcommand{\CameraReadyIdleRows}{%
Original & 20 & 46 [0--104.75] & 12 & 4 \\
Forbidden & 10 & 90 [0--116] & 8 & 4 \\
Fixed & 29 & 85 [49--556] & 0 & 0 \\
\texttt{prec15} & 18 & 31.5 [0--99] & 12 & 4 \\
\texttt{prec25} & 12 & 0 [0--37.75] & 7 & 0 \\
\midrule
All-126 & 89 & 47 [0--116] & 39 & 12 \\
}

\makebox[\columnwidth][c]{%
\begin{tabular}{@{}lrccc@{}}
\toprule
Family & Feasible & \shortstack{$I_\Sigma$\\median [IQR]} & \shortstack{Zero\\range} & \shortstack{Zero range,\\positive idle} \\
\midrule
\CameraReadyIdleRows
\bottomrule
\end{tabular}
}%
\end{table}

\begin{table}[t]
\caption{All-126 Solved Instances and Total Time. 7200-s Limit; One Run per Configuration--Instance.}
\label{tab:cross-paradigm}
\centering
\scriptsize
\setlength{\tabcolsep}{1.3pt}
\begin{tabular}{@{}l@{\hspace{6pt}}c@{\hspace{5pt}}c@{\hspace{5pt}}cc@{}}
\toprule
Method & Solved & Timeouts &
  \multicolumn{2}{c}{Total time (s)\,$\downarrow$} \\
\cmidrule(lr){4-5}
& & & Median [IQR] & PAR-2 \\
\midrule
\textsc{Incremental SAT} & 126 & 0 &
  \textbf{0.227 [0.119--0.400]} & 1.266 \\
\textsc{Non-Incremental SAT} & 126 & 0 & 0.312 [0.126--0.506] &
  \textbf{0.547} \\
UWrMaxSAT & 126 & 0 & 0.432 [0.192--0.668] & 0.684 \\
Gurobi MIP & 126 & 0 & 0.521 [0.096--9.614] & 29.755 \\
CPLEX MIP & 123 & 3 & 1.611 [0.358--15.111] &
  435.815 \\
CP Optimizer & 115 & 11 & 0.110 [0.051--0.684] &
  1301.351 \\
\bottomrule
\end{tabular}
\end{table}

On All-126, \textsc{Incremental SAT} has the lowest median among methods solving all instances (0.227~s).  \textsc{Non-Incremental SAT} has the lowest PAR-2 because one incremental run takes 104.873~s. For RQ4, the three SAT and MaxSAT methods and Gurobi solve all 126 instances. CPLEX MIP and CP Optimizer solve 123 and 115 instances, respectively.  CP Optimizer has the lowest median (0.110~s) but times out on 11 instances.  On the 123 instances solved by both MIP solvers, Gurobi has lower total time in 112 cases.  The median per-instance ratio of CPLEX MIP total time to Gurobi total time is 2.78.  Results reflect single-run timing; RQ1 and RQ2 isolate domain and precedence components, and Stress-100 controls for precedence-density effects.

\section{Conclusion}
\label{sec:conclusion}

Our compact SAT and MaxSAT model combines solution-preserving filtering, exact suffix sharing, and idle-time range balancing. Under the common $\Delta_I$ objective, its median paired reductions are 40.3\% in clauses and 55.9\% in peak memory relative to \mbox{\textsc{ORG-MaxSAT}}. Domain reduction alone removes 16.2\% of clauses. SparseSuffix gives a smaller additional reduction of 0.5\%--1.0\% on official precedence instances, reaching 5.5\% under \mbox{\textsc{Closure-}$E^\star$} at \texttt{prec60} in the controlled derived set. Its total-time changes are small and mixed. Both filters reach the same feasible fixpoints; \mbox{\textsc{Filter-}$E^\star$} uses fewer passes on 118/140 cases and never more. In the original one-run comparison, all three SAT and MaxSAT methods solve All-126 with lower median times than Gurobi, the only commercial method to solve every instance. Incremental SAT has the lowest median among methods solving all instances. A zero range can coexist with positive aggregate idle, so the two measures describe complementary aspects of schedule quality.

The largest tested instances contain 302 meetings and 78 participants. Stress-100 varies precedence density while keeping the numbers of meetings, participants, and slots unchanged. Larger events may require more time for graph construction and participant matching. Future work will study larger instances, rooms with different capacities or equipment, and objectives that minimize $\Delta_I$ first and then $I_\Sigma$.

\end{document}